\documentclass[
    prd, twocolumn, superscriptaddress, nofootinbib, amsmath, amssymb,
    aps, floatfix, preprintnumbers
]{revtex4-2}
\usepackage[utf8]{inputenc}
\usepackage[dvips]{graphicx}
\usepackage{booktabs}
\usepackage[dvipsnames]{xcolor}
\definecolor{CiteBlue}{RGB}{45,52,151}
\usepackage[
    colorlinks=true,
    linkcolor=CiteBlue,
    urlcolor=CiteBlue,
    citecolor=CiteBlue
]{hyperref}
\usepackage[version=4]{mhchem}
\usepackage{aas_macros}
\usepackage{multirow}

\usepackage[capitalise]{cleveref}
\usepackage{siunitx}

\newcommand{\refcite}[1]{Ref.~\cite{#1}}
\newcommand{\refscite}[1]{Refs.~\cite{#1}}

\usepackage{bm}
\newcommand{\bb}[1]{\bm{\mathrm{#1}}}
\newcommand{\du}{\mathrm{d}}
\newcommand{\dd}{\,\du}
\newcommand{\dm}{\mathrm{DM}}
\newcommand{\med}{\phi}
\renewcommand{\Im}{\operatorname{Im}}
\newcommand{\decay}{-}
\newcommand{\rise}{+}
\newcommand{\obs}{\mathrm{obs}}
\newcommand{\bg}{\mathrm{bg}}

\begin{document}

\title{First direct search for light dark matter interactions in a transition-edge sensor}
\preprint{MIT-CTP/5879, DESY-25-086}

\author{Christina Schwemmbauer}
\affiliation{Deutsches Elektronen-Synchrotron DESY, Notkestr. 85, 22607 Hamburg, Germany}

\author{Guy Daniel Hadas}
\affiliation{Racah Institute of Physics, Hebrew University of Jerusalem, Jerusalem 91904, Israel}

\author{Yonit Hochberg}
\affiliation{Racah Institute of Physics, Hebrew University of Jerusalem, Jerusalem 91904, Israel}
\affiliation{Laboratory for Elementary Particle Physics, Cornell University, Ithaca, NY 14853, USA}

\author{Katharina-Sophie Isleif}
\affiliation{Helmut-Schmidt-Universität, 22043 Hamburg, Germany}

\author{Friederike Januschek}
\affiliation{Deutsches Elektronen-Synchrotron DESY, Notkestr. 85, 22607 Hamburg, Germany}

\author{Benjamin V. Lehmann}
\affiliation{Center for Theoretical Physics -- a Leinweber Institute, Massachusetts Institute of Technology, Cambridge, MA 02139, USA}

\author{Axel Lindner}
\affiliation{Deutsches Elektronen-Synchrotron DESY, Notkestr. 85, 22607 Hamburg, Germany}

\author{Adriana E. Lita}
\affiliation{National Institute of Standards and Technology, Boulder, Colorado, USA}

\author{Manuel Meyer}
\affiliation{CP3-origins, Department of Physics, Chemistry and Pharmacy, University of Southern Denmark, Campusvej 55, 5230 Odense, Denmark}

\author{Gulden Othman}
\affiliation{Helmut-Schmidt-Universität, 22043 Hamburg, Germany}

\author{Elmeri Rivasto}
\affiliation{CP3-origins, Department of Physics, Chemistry and Pharmacy, University of Southern Denmark, Campusvej 55, 5230 Odense, Denmark}

\author{Jos\'e Alejandro Rubiera Gimeno}
\affiliation{Helmut-Schmidt-Universität, 22043 Hamburg, Germany}

\date\today

\begin{abstract}\ignorespaces{}
We propose the use of transition-edge sensor (TES) single-photon detectors as a simultaneous target and sensor for direct dark matter searches, and report results from the first search of this kind. We perform a \qty{489}{\hour} science run with a TES device optimized for the detection of \qty{1064}{\nano\meter} photons, with a mass of ${\sim}\qty{0.2}{\nano\gram}$ and an energy threshold of ${\sim}\qty{0.3}{\electronvolt}$, and set new limits on dark matter interactions with both electrons and nucleons for dark matter with mass below the MeV scale. With their excellent energy resolution, TESs enable search strategies that are complementary to recent results from superconducting nanowire single-photon detectors and kinetic inductance detectors. We show that next-generation TES arrays hold promise to probe new regions of light dark matter parameter space.
\end{abstract}

\maketitle

\section{Introduction}\label{sec:intro}

Dark matter (DM) makes up the majority of matter in our universe, yet its identity remains unknown. For decades, theoretical work focused on weak-scale DM candidates guided the majority of experimental searches aimed at detecting such DM candidates. The nondetection of weak-scale DM particles at a variety of dedicated experiments~\cite{Feng:2010gw, Bertone:2018krk, Akerib:2022ort, ParticleDataGroup:2024cfk, Bergstrom:2000pn} has ushered in a new era of exploration in the DM community, with new emphasis on candidates with sub-GeV masses (see \refcite{Asadi:2022njl} for a review). In turn, novel experimental techniques have emerged to search for such particles in the laboratory~\cite{Essig:2011nj,Graham:2012su,Essig:2015cda,Hochberg:2015pha,Hochberg:2015fth,Hochberg:2019cyy,Hochberg:2021ymx,Hochberg:2021yud,Derenzo:2016fse,Hochberg:2016ntt,Hochberg:2017wce,Cavoto:2017otc,Kurinsky:2019pgb,Blanco:2019lrf,Griffin:2020lgd,Simchony:2024kcn,Essig:2022dfa,Das:2022srn,Das:2024jdz}. Among these, superconducting targets~\cite{Hochberg:2015pha,Hochberg:2015fth} stand out due to their potential sensitivity to energy deposits as low as $\mathcal{O}(\qty{}{\milli\electronvolt})$. This would enable the detection of DM as light as \qty{1}{\kilo\electronvolt}, below which cosmological constraints robustly exclude fermionic DM. Such detectors would also be sensitive to the absorption of bosonic DM with masses as low as \qty{1}{\milli\electronvolt}~\cite{Hochberg:2016ajh}. The status of DM searches has thus motivated new improvements in the design of superconducting detectors.

Indeed, rapid developments in quantum sensing are quickly enabling groundbreaking progress in DM research.\ \refscite{Hochberg:2019cyy,Hochberg:2021yud} used a prototype superconducting nanowire single-photon detector (SNSPD) as both the sensor to measure DM interactions and the target mass with which the DM interacts, placing world-leading limits on light DM with mass beneath the MeV scale. These first-generation SNSPD limits were then surpassed by the second-generation QROCODILE experiment~\cite{QROCODILE:2024zmg}, with an experimental setup specifically developed for the detection of light DM. SNSPDs are, however, not the only superconducting sensors that can be used in this manner.  \refcite{Gao:2024irf} previously proposed the use of kinetic inductance detectors (KIDs) as DM detectors. In this work, we demonstrate the use of transition-edge sensors (TESs) as the superconducting target and sensor for light DM detection, as we proposed in \refscite{Schwemmbauer:2024jel, Schwemmbauer:2024rcr}. TESs have much greater energy resolution than SNSPDs, while still offering favorable noise characteristics. Advances in TES design thus enable new complementary methods of probing light DM parameter space. 

TESs are broadly used in science and industry. For example, TES arrays are used in the cameras of large telescopes~\cite{Holland:2013ng, 2009AIPC.1185..475C} and in quantum information technologies (see \textit{e.g.} \refcite{2011PhRvA..84f0301G}). TESs are also widely used in DM detection experiments such as CRESST~\cite{Gascon:2010zz} and SuperCDMS~\cite{SuperCDMS:2022kse}. Currently, however, they are typically used as sensors coupled to a separate target mass, as in the TESSERACT experiment~\cite{TESSERACT:2025tfw}: the DM interacts with the target material, and a portion of the deposited energy is later detected by the TES. Early proposals for superconducting light DM searches similarly suggested the use of TESs as the sensor that instruments a large bulk target made of a material such as aluminum~\cite{Hochberg:2015pha,Hochberg:2015fth}, using a detection philosophy similar to that of \textit{e.g.} the semiconductor targets of SuperCDMS.

Here, we take an alternative complementary path: for the first time, we demonstrate the use of a TES as \textit{simultaneously} the target with which the DM interacts and as the sensor that detects this interaction. This maximizes the sensitivity to very light DM candidates, where detection is bottlenecked by the effective detector threshold. Since TESs are mainly sensitive to the amount of energy deposited, the scattering of a DM particle in the TES causes the device to respond in the same manner as to a photon depositing a similar amount of energy. In particular, as long as the energy deposited surpasses the noise level while not driving the TES into its normal conducting region, the calorimetric characteristics of the latter can be exploited. The pulse shape of the output signal then only depends on the particle’s energy~\cite{Hochberg:2021ymx,Griffin:2024cew}, and not on other the properties of the incoming particle, whether a photon or a DM particle.

In this work, we report new measurements from an existing TES device intended for use in the Any Light Particle Search II (ALPS~II) experiment~\cite{Gimeno:2023nfr}, optimized for the detection of photons originating from axions and axion-like particles. We use our measurements to place new limits on DM with sub-MeV masses that interacts with electrons or with nuclei in the TES. Our work demonstrates the potential of this approach, and highlights the prospects to harness advances in quantum sensing to push fundamental physics forward.

This paper is organized as follows. \Cref{sec:exp} describes the experimental setup, including calibration and backgrounds. \Cref{sec:run} outlines our DM science run and analysis, and \cref{sec:DM} describes our computation of DM interaction rates in the device. Our results are given in \cref{sec:res}, and we conclude with an outlook for future experiments in \cref{sec:out}. Throughout this work, we use natural units, with $c = \hbar = k_{\mathrm{B}} = 1$.

\section{Experimental setup}
\label{sec:exp}

\subsection{TES device}
\label{sec:tes}

TESs are superconducting microcalorimeters operated at cryogenic temperatures of $\mathcal{O}(\qty{10}{\milli\kelvin})$~\cite{2005cpd..book...63I}. The superconductor is thermally connected to a cold bath and voltage-biased to a working point slightly above the critical temperature $T_{\mathrm{C}}$ between the superconducting and normal conducting state, that is, on the edge of the transition. When biased at the working point, an energy deposit above the noise threshold in the TES will lead to a small temperature change $\Delta T$, leading to a large increase in resistance $\Delta R$ due to the steep rise of the resistance-temperature curve. The change in resistance leads to a changing current in the circuit. The changing current is measured by the changing magnetic flux through an inductively-coupled coil, which is in turn read out by a Superconducting Quantum Interference Device (SQUID). The SQUID output shows a voltage pulse, proportional to the energy originally deposited in the TES by the incident particle. The shape of the pulse above the baseline voltage $V_0$ can be described by the following expression based on small signal theory~\cite{2005cpd..book...63I}:
\begin{equation}
    \label{eq:template}
    V(t) - V_0 = \begin{cases}    
    A\left[
        e^{-(t-t_0)/\tau_{\rise}}
        - e^{-(t-t_0)/\tau_{\decay}}
    \right] & t \geq t_0\\
    0 & t < t_0,
    \end{cases}
\end{equation}
for appropriate choices of start time $t_0$, amplitude $A$, rise time $\tau_{\rise}$, and decay time $\tau_{\decay}$. The pulses can generally be fit well in the frequency domain, with a template given by the Fourier transform of \cref{eq:template}. In practice, this fit is performed numerically using the fast Fourier transform (FFT; see \refcite{RubieraGimeno:2024} for further details of the FFT fitting procedure). Previous work using a similar detector setup has demonstrated a linear relationship between deposited energy and voltage pulse height in the readout for photons with energies between \qty{1}{\electronvolt} and \qty{3}{\electronvolt}~\cite{Bastidon:2016, Dreyling-Eschweiler:2015pja}.

\begin{figure}\centering
    \includegraphics{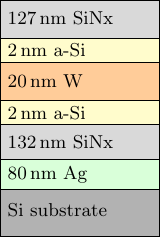}
    \hfill
    \includegraphics[width=5.4cm]{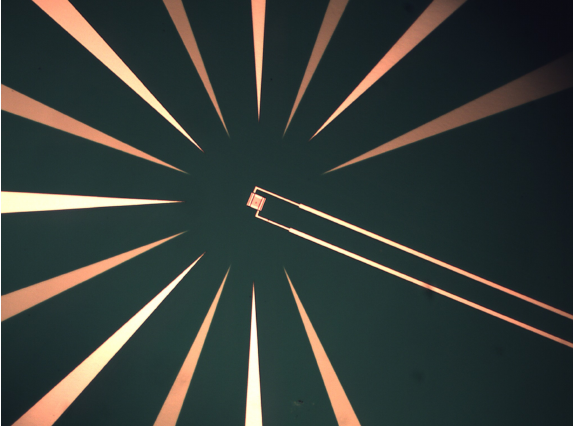}
    \caption{
        Prototype TES detector.
        \textit{Left:}~Schematic showing different layers of the optical TES stack with the active tungsten (W) layer placed in between dielectric amorphous Si-layers and on a reflective Ag mirror and Si substrate. This layering optimizes the absorption at \qty{1064}{\nano\meter}.
        \textit{Right:}~Closeup of the TES sensor on the silicon substrate, provided by NIST; the fins around the square chip are for aligning purposes only (taken from \refcite{Shah:2023kef}).
    }
    \label{fig:tes-device}
\end{figure}

The TES module used in this work, provided by NIST and shown in the right panel of \cref{fig:tes-device}, is the device to be used for 1064~nm photon counting at the ALPS~II experiment. It consists of a tungsten TES with an area of $\qty{25}{\micro\meter}\times\qty{25}{\micro\meter}$, a thickness of \qty{20}{\nano\meter}, and a mass of ${\sim}\qty{0.2}{\nano\gram}$ embedded in an optical stack~\cite{2010SPIE.7681E..0DL}, shown schematically in the left panel of \cref{fig:tes-device}. The optical stack is designed to optimize the absorption of light at \qty{1064}{\nano\meter} wavelength, corresponding to an energy of \qty{1.165}{\electronvolt}. This is the wavelength of the laser used for axion-like-particle production in ALPS~II.

The setup and procedure for performing a direct DM search with our TES builds on previous work done for ALPS~II. For the development of the ALPS~II analysis procedure, the intrinsic background (\textit{e.g.} natural radioactivity) was recorded in the TES's environment, the interior of a dilution refrigerator approximately two floors underground in the HERA West building at DESY. The data from these intrinsic background measurements are free from outside light or heat sources introduced \textit{e.g.} by optical fibers coupled to the TES, offering appropriate conditions for a direct DM search as well. These measurements and other possible background sources are discussed in detail in \refcite{Shah:2023kef}. To perform direct DM searches with this setup, a data acquisition~(DAQ) trigger level that is as low as technically possible needs to be determined in order to reduce the effective energy threshold of the TES for photon-like pulses to sub-eV levels while maintaining a sufficient signal-to-noise ratio. These results are then directly applicable to a search for photon-like events from the scattering or absorption of light DM particles in the TES.

\subsection{TES configuration and calibration}
\label{sec:config}

Our TES module consists of two separate TES chips wire-bonded to a single-stage dc-SQUID series array chip \cite{4277368, 2018JLTP..193.1243S} for biasing and readout via cryogenic cables and specialized readout electronics. In this work, we use data recorded with only one of the TES chips. The TES modules' packaging and SQUIDs were provided by PTB, Germany. The TES is enclosed by an aluminum can for noise reduction. The aluminum can and TES module are attached to the lowest stage of a BlueFors dilution refrigerator, which enables cooling below \qty{25}{\milli\kelvin}. By applying an appropriate bias current, the TES chips are operated at 20\%--30\% of the normal-state resistance $R_{\mathrm N}$. Choosing a proper TES working point for the measurements is especially critical as the working point influences the pulse integral, energy resolution, and noise of the recorded signals~\cite{Shah:2023kef}, impacting the detector's sensitivity to low energies. While higher working points closer to the normal conducting phase exhibit lower noise, the used lower working points offer better energy resolution. We trigger on signals from the readout via a digitizer using a sampling rate of $\qty{50}{\mega\hertz}$.

For intrinsic background measurements, optical fibers used for photon measurements are disconnected from the TES, removed from the detection volume inside the aluminum can, and placed on the flange above. While the fibers are therefore mostly isolated from the TES' operational volume, a small gap between the flange and aluminum allows light from the fiber tips to scatter into the detection volume for calibration measurements prior to dedicated background measurements. 

For direct DM searches, the SQUID settings, TES working points, and DAQ trigger levels are optimized to guarantee a high dynamic range as well as low noise while still providing high energy resolution. Before a measurement, single-photon samples with attenuated laser sources are recorded to determine the signal-to-noise ratio for photon pulses in the energy region of interest. To choose an appropriate trigger level for low energy events, a short continuous timeseries is recorded from the readout without any trigger. This data sample predominantly consists of the current noise at the TES output, measured with the SQUID sensor and readout electronics. We will refer to this as 'noise' in the remainder of this work. The sample is then analyzed to find a suitable trigger level, i.e., an energy threshold which is as low as possible without reaching significantly into the noise baseline, where a high trigger rate could potentially lead to deadtime. We select a trigger level corresponding to a trigger rate of less than \qty{2}{\hertz}. Consequently, in order to trigger on small energy deposits, the background levels of the system must be reduced as much as possible. In the post-trigger analysis, interesting signals are then selected through a pulse shape analysis similar to that in \refscite{Shah:2021wsp,RubieraGimeno:2024}, where the template of \cref{eq:template} is fitted to the signal in the frequency domain.

\subsection{Backgrounds}\label{sec:bg}

As is evident from dedicated intrinsic background measurements for ALPS~II~\cite{Shah:2021wsp}, non-negligible backgrounds are already present at energies above \qty{1}{\electronvolt}. These include ambient radioactivity from materials that are part of the setup or energy depositions by cosmic rays (\textit{e.g.} in the silicon substrate). For example, it is likely that the zirconium dioxide fiber sleeves surrounding the TES chips include radionuclides that could produce spurious signals in the TES~\cite{Bastidon:2016}. This is also supported by dedicated simulations matching the expectations from background measurements~\cite{RubieraGimeno:2024}. Subsequent analysis procedures can reduce the intrinsic backgrounds at \qty{1.165}{\electronvolt} to below $\qty{6.9e-6}{\per\second}$ while maintaining up to 90\% acceptance for \qty{1064}{\nano\meter} signal photons~\cite{Shah:2021wsp,Meyer:2023ffd}. The signal acceptance is expected to be lower for lower energy pulses due to a decrease in the signal-to-noise ratio. Due to the lowered DAQ trigger threshold compared to ALPS~II studies, the majority of background counts in our dataset are expected to originate from noise of the detection system. 

The pulse shape analysis (thoroughly discussed in \refcite{RubieraGimeno:2024}), uses fitting functions for expected photon-signal shapes and implements cuts on parameters such as rise time $\tau_\rise$ and decay time $\tau_\decay$. This can be adapted for lower signal energies as well, according to the applied trigger. Therefore, we employ the pulse shape analysis to mitigate the impact of backgrounds on our DM constraints by eliminating pulses that are clearly distinct from photon-like signals. A similar study of photon-like dark counts in a TES has been performed for part of our investigated energy range using a principal component analysis (PCA) in \refcite{Manenti:2024etv}. Further mitigation of backgrounds may be achieved using machine learning techniques, as in \refcite{Meyer:2023ffd}, and we defer this to future work.

\subsection{Calibration and pulse shape parameters}

\begin{figure}
    \centering
    \includegraphics[width=\linewidth]{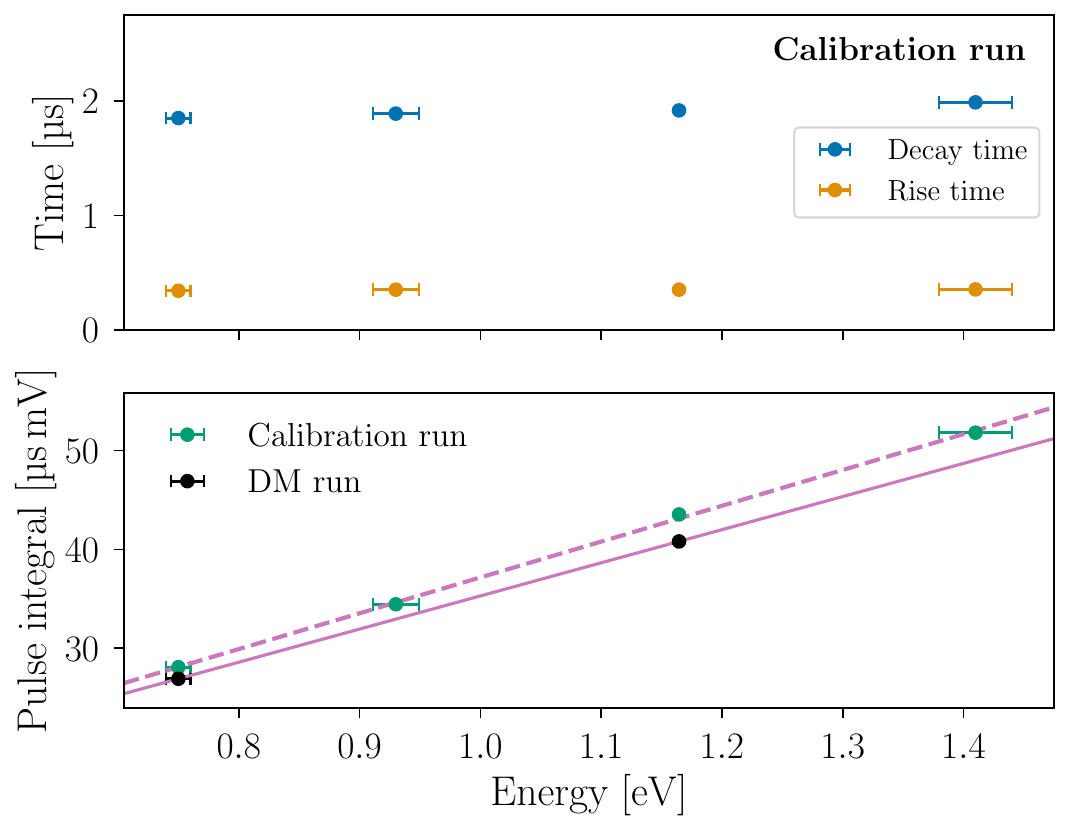}
    \caption{Calibration measurements. \textit{Top:} Mean fitted rise time $\tau_{\rise}$ and decay time $\tau_{\decay}$ as defined by \cref{eq:template}. The time uncertainties are not visible at this scale. The uncertainties of the energy represent the spectral width of the laser diodes, as measured with a spectrometer. The rise and decay time of the calibration photons are mostly constant over this energy range.\ \textit{Bottom:} Mean pulse integral from the fitting function for different energies for our calibration run (green) and DM run (black). The uncertainties on the pulse integral are not visible at this scale. The relationship between the pulse integral and energy of the calibration photons follows an affine relation as indicated by the dashed purple line. In our analysis, we assume that a similar relation also holds for the DM run, indicated by the solid purple line. Environmental changes between different cooldowns can cause differences in the calibration relations.
    }
\label{fig:calib}
\end{figure}

The aforementioned linear relationship between deposited energy and voltage pulse height has only been shown for energies ${>}\qty{1}{\eV}$ with a similar setup. Dedicated simulations for a larger energy range show the same behavior for the pulse integral, which stays linear up to higher energies compared to the pulse height~\cite{RubieraGimeno:2024}. Therefore, the pulse integral is used as a calibration parameter. We perform several calibration measurements to verify this relationship for our current setup, to explore the energy dependence below the \qty{1}{\electronvolt} scale, and to study the dependence of pulse rise and decay times on the amount of energy deposited. In our measurements, we expose the device to heavily attenuated continuous-wave laser light from laser diodes with wavelengths of \qty{1640}{\nano\meter}, \qty{1310}{\nano\meter}, \qty{1064}{\nano\meter}, and \qty{880}{\nano\meter}, corresponding to energies of \qty{0.756}{\electronvolt}, \qty{0.947}{\electronvolt}, \qty{1.165}{\electronvolt}, and \qty{1.409}{\electronvolt}. Testing for lower energies is not possible with this setup due to attenuation effects: since the fiber is optimized for a wavelength of \qty{1064}{\nano\meter} and curled within the setup, higher wavelengths are not transmitted~\cite{Jay:2010}. Simulations taking the fiber and curling radii into account show a sharp cutoff of transmitted light below ${\sim}\qty{0.7}{\electronvolt}$ (above ${\sim}\qty{1800}{\nano\meter}$)~\cite{RubieraGimeno:2025bhr}.

We performed a first calibration measurement in March 2024, with the results shown in \cref{fig:calib}. The top panel of \cref{fig:calib} shows that, as expected, the rise and decay times are independent of the pulse energy over the range of measurement. The calibration run data in the bottom panel of \cref{fig:calib} verifies a nearly-linear relationship between incident photon energy and the integral $\mathcal I$ of the associated pulse, maintained to energies at least as low as \qty{0.756}{\electronvolt}. We fit an affine relation, \textit{i.e.,} we set $E - E_0 \propto \mathcal I$. The results indicate a non-zero intercept of the calibration curve, originating from noise baseline fluctuations. This effect has been observed in similar TES setups as well (see \refcite{Manenti:2024etv}). In our current configuration, testing the TES response at lower energies is not possible due to aforementioned attenuation effects. We thus extrapolate the observed behavior down to lower energies.

During the same cooldown as the DM science run, we performed an additional set of calibration measurements using \qty{1064}{\nano\meter} and \qty{1640}{\nano\meter} lasers, corresponding to energies of \qty{1.165}{\electronvolt} and \qty{0.756}{\electronvolt}. These measurements were performed directly before the DM science run. This additional calibration is necessary since the detection module is very sensitive to environmental changes: each new cooldown can lead to a variation in pulse shape, even for the same energies. We use these two measurements, indicated by the black points in the bottom panel of \cref{fig:calib}, to establish the parameters of the affine relationship between deposited energy and pulse integral assumed in our DM science run (\textit{i.e.}, the solid purple line).

Based on the stability of the rise and decay time response, as well as the near-linearity of the pulse integral (as holds for the pulse amplitude $A$), we simulate datasets of photon-like energy deposits from \qtyrange{0.1}{3.0}{\electronvolt} using the framework described in \refcite{RubieraGimeno:2022pjx,RubieraGimeno:2024}. We use these simulated pulses as a guideline to determine cuts for our DM analysis. The stable rise and decay time are two of the main indicators of the photon-like shape of a pulse. Therefore, cuts on these parameters help to isolate photon-like pulses from the majority of the background measured with a low trigger level. We use the simulated dataset to determine the effective acceptance of our experiment when different cuts are enforced. The affine relationship between the energy and the pulse integral is then used to determine the energy spectrum of events that survive these cuts.

\section{Data and analysis}\label{sec:run}
\begin{figure}
    \centering
    \includegraphics[width=0.5\textwidth]{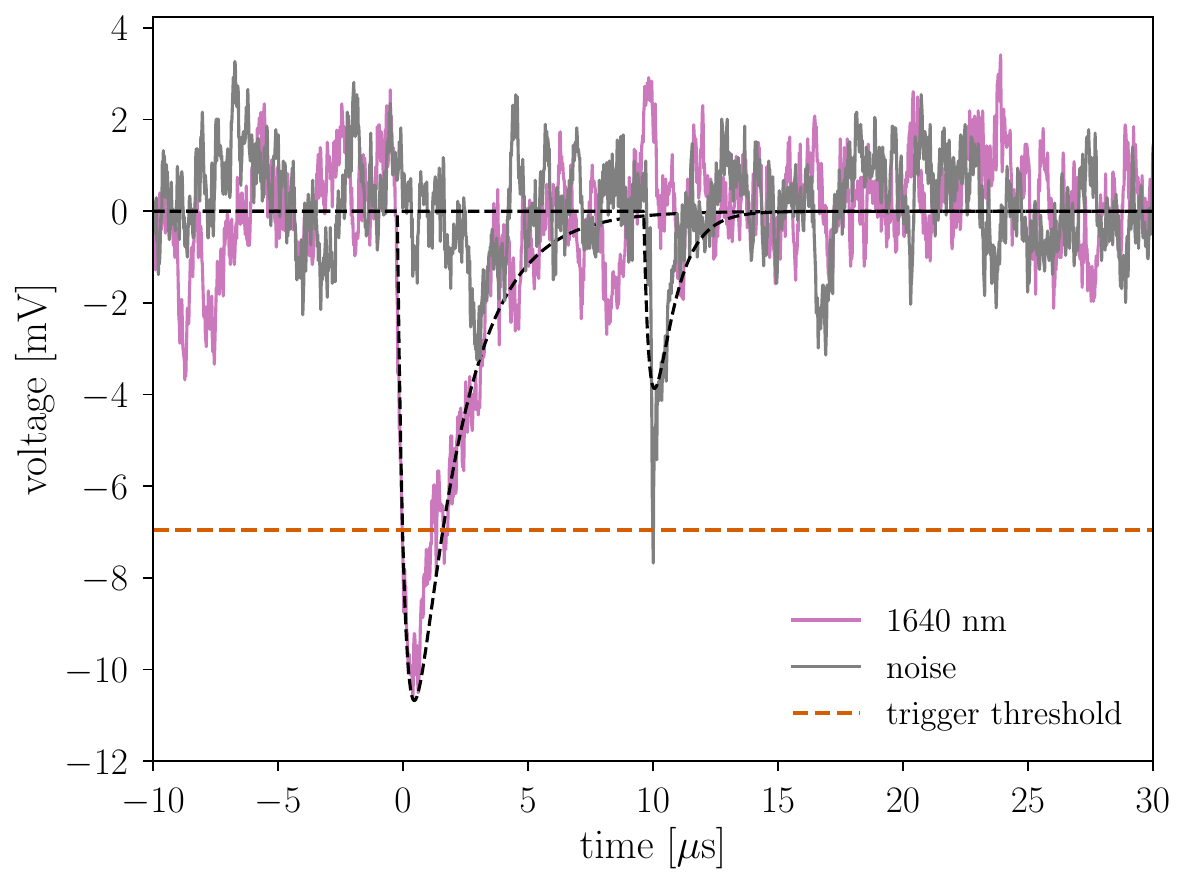}
    \caption{Example pulses. Example pulses measured with the TES system. The purple curve shows timeseries measurements for a \qty{1640}{\nano\meter} calibration pulse (DM run), while the gray curve represents a pulse yielding a trigger from noise recorded as part of the DM search measurement and removed before the cut analysis. The dashed black line represents the FFT fits and the red dashed line shows the applied trigger level. A mismatch between the pulse height of the fit and the minimum of the timeseries itself is clearly visible for the noise spike. But for a single sample surpassing the threshold, our pipeline would not have triggered on this pulse.}
    \label{fig:example-pulses}
\end{figure}

 \begin{figure*}
        \centering
        \includegraphics[width=0.48\textwidth]{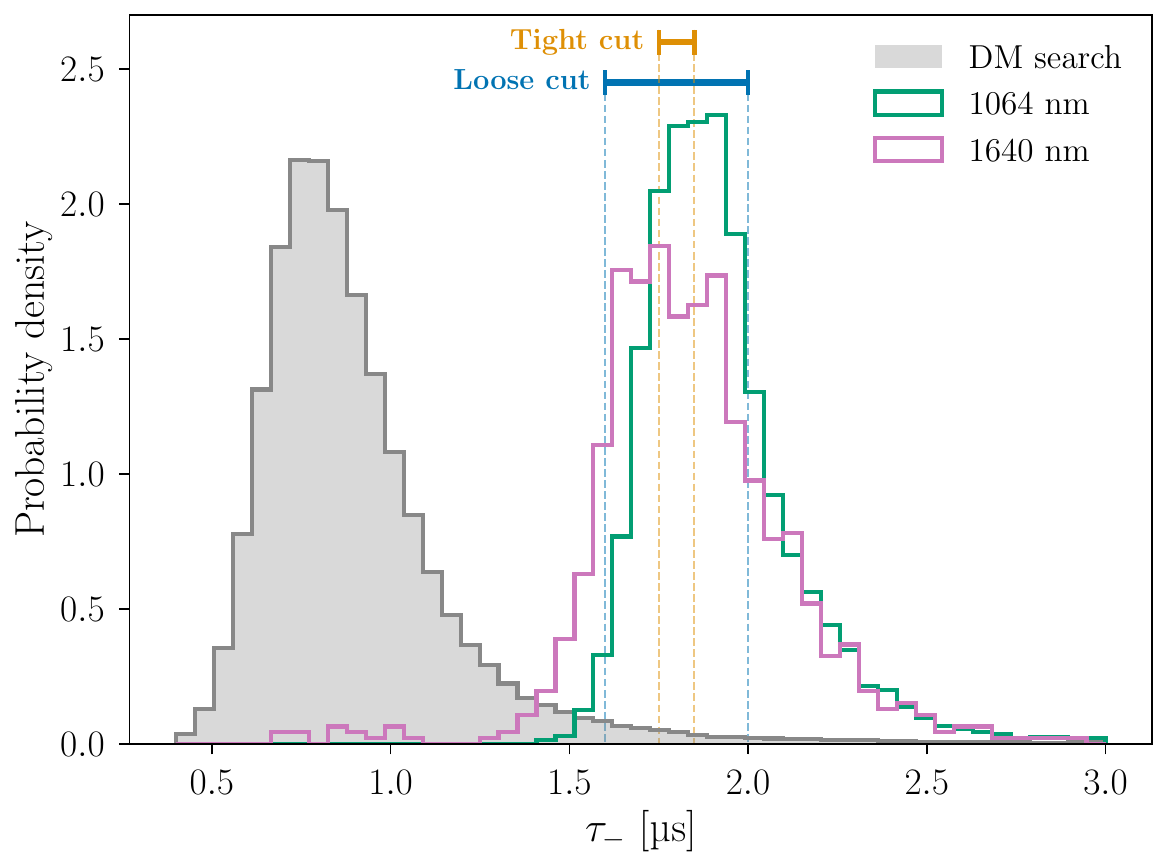}
        \hfill
        \includegraphics[width = 0.48\textwidth]{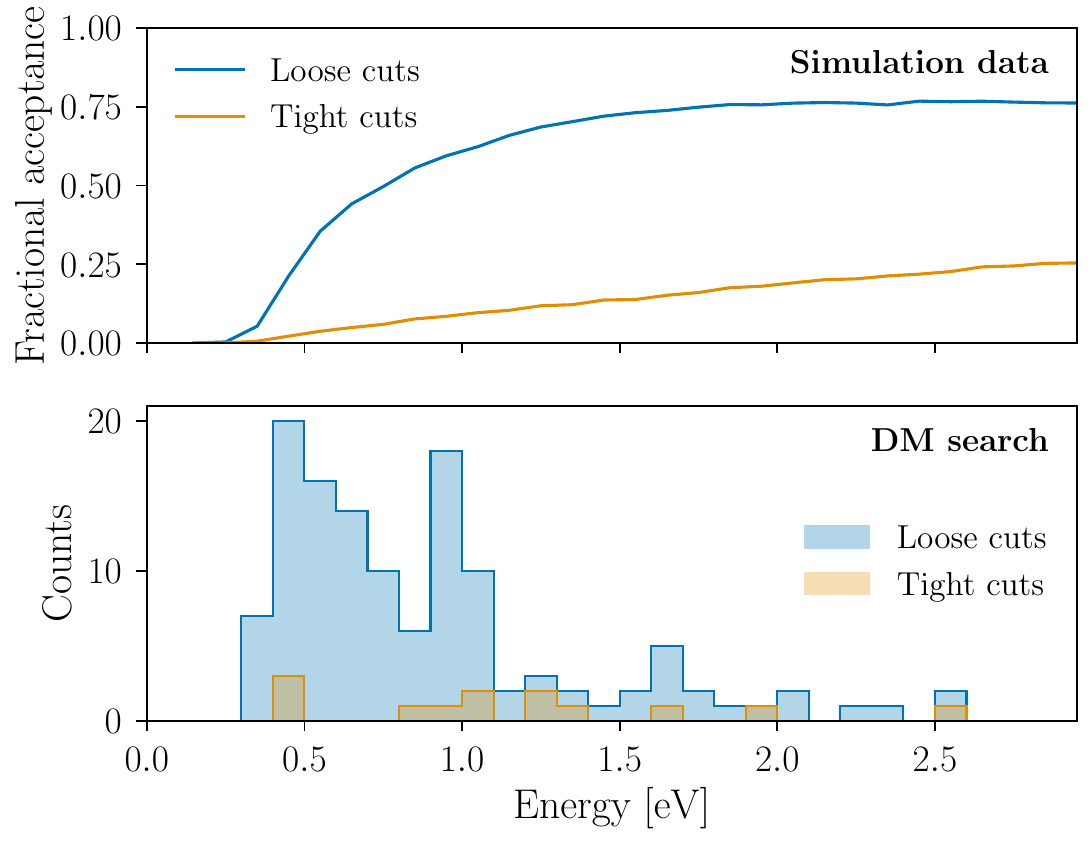}
        \caption{DM science run. \textit{Left:} Distribution of FFT fitting parameter $\tau_\decay$ of the direct DM search measurement (gray) and the two measured calibration wavelengths (green and magenta) for reference. While the lasers' distributions overlap with each other to a large extent, much of the DM search measurement distribution lies outside of that region due to small decay times of fast noise spikes. Rigorous cuts can be applied based on this distinction, which also holds for photon-like signals of a broader energy range.\ \textit{Right top:} Total acceptance $\mathcal A$ in simulation data for the loose and tight cuts (see \cref{tab:cuts}).\ \textit{Right bottom:} Energy spectrum measured over \qty{489}{\hour} and calibrated with the FFT pulse integral parameter based on the loose and tight cuts. 
        The bin size approximates the energy resolution of ${\sim}\qty{0.1}{\eV}$ at ${\sim}\qty{0.76}{\electronvolt}$.
        }
    \label{fig:counts-and-cuts}
\end{figure*}

We now describe our DM science run which took place in April and May of 2024. 
The analysis for the DM science run was performed in the frequency domain based on pulse shapes predicted by small-signal theory~\cite{2005cpd..book...63I}. We perform the frequency-domain analysis using the \texttt{TESPASS} framework, described in detail in \refscite{RubieraGimeno:2024, Gimeno:2023nfr}. The following results are based on a \qty{489}{\hour} intrinsic background measurement (with no fiber) with a trigger level and setup following the procedures described in \cref{sec:config}.

\subsection{Pulse identification and cuts}

Prior to the fitting analysis and cuts, we use the pulse-finding algorithm of \refcite{RubieraGimeno:2024} to reject pulses with shapes very different from a photon-like pulse, such as many of those arising from noise. We perform this selection using signal deconvolution in combination with a low-pass filter with a cutoff frequency at \qty{0.5}{\mega\hertz} and Gaussian kernel for the deconvolved signals. For each pulse, we fit the pulse parameters in the frequency domain, identifying the starting time, rise time, decay time, and pulse integral. Some fast noise signals, barely passing the trigger level, still survive the pulse finder and exhibit a pulse shape deviating from the expected one.\footnote{This mismatch between a fast noise spike passing the trigger with only one or two data samples in the timeline and signal pulse shape is not always properly represented in the $\chi^2_\mathrm{red}$ of the frequency domain fit. The fit then represents a fluctuation in the noise baseline, which would not have passed the trigger otherwise, where the dominating uncertainties in the frequency domain lead to good $\chi^2_\mathrm{red}$ values. Therefore, these fast spikes need to be isolated differently.} Hence, pulses are only further considered when the deviation ratio between the minimum of the measured signal and the fitted pulse-height, $\mathcal{R}_\mathrm{dev}$, falls within the $3\sigma$ region of the distribution of that ratio determined for the \qty{1640}{\nano\meter} calibration pulses. \Cref{fig:example-pulses} shows one of the \qty{1640}{\nano\meter} calibration pulses compared to a fast noise spike from the same measurement. Here, the pulse height found with the fitting function exhibits a distinct mismatch to the minimum of the pulse timeseries itself. This noise fluctuation only surpasses the trigger level for one sample of the timeseries, and is therefore removed from the dataset based on the aforementioned selection criterion.

We then impose cuts on the rise time, decay time, and reduced $\chi^2$ between the data and the pulse-shape template. We perform our entire analysis twice with two different sets of cuts, described in \cref{tab:cuts}. These cuts are based on the distribution of fit parameters of the calibration data recorded just before the DM science run, along with simulated data sets of photon-like pulses over an energy range of \qtyrange{0.1}{3.0}{\electronvolt}, also simulated using the {\tt{TESPASS}} framework. The cuts reject pulses with fitting parameters outside of the expected region for photon-like pulses. For this study, the specific cuts were chosen based on the simulated signal distributions to approximate ${\gtrsim}50\%$ acceptance at ${\sim}\qty{1}{\eV}$ for the less restrictive cuts and ${\sim}10\%$ acceptance for the more restrictive cuts. The less restrictive ``Loose'' cuts allow for a higher acceptance at the cost of a higher background rate, while the set of ``Tight'' cuts gives reduced acceptance but substantially attenuates the background rate.

\begin{table}
    \centering
    \begin{equation*}
        \begin{array}{lccc}
            \toprule
                & \textnormal{\textbf{Parameter}}
                    & \textnormal{\textbf{Loose}}
                    & \textnormal{\textbf{Tight}}
            \\\midrule
                \textnormal{\textbf{Cuts}}
                    & \tau_\rise~[\qty{}{\micro\second}]
                        & (0.3, 0.4) & (0.325, 0.375) \\
                    & \tau_\decay~[\qty{}{\micro\second}]
                        & (1.6, 2.0) & (1.75, 1.85) \\ 
                    & \chi^2_\text{red}
                        & (0.9, 1.1)& (0.95, 1.05) \\ 
                    & t_0~[\qty{}{\micro\second}]
                        & (-0.5, 0.5) & (-0.5, 0.5)\\
                    & \mathcal{R}_\mathrm{dev}
                        & (0.73, 1.10) & (0.73, 1.10)
            \\\midrule
                \textnormal{\textbf{Results}}
                    & \textnormal{Event rate}~[\qty{}{\hertz}]
                        & \num{7.2e-5} & \num{7.4e-6} \\ 
                    & \textnormal{Survival}~[\qty{}{\percent}]
                        & 0.107 & 0.011 \\ 
            \bottomrule
        \end{array}
    \end{equation*}
    \caption{Parameter cuts. Summary of loose and tight cuts applied to the DM search data (frequency domain analysis), with  ranges in parentheses denoting lower and upper cut values. We also show the resulting event rate and percentage of surviving triggers for each set of cuts.
    }
    \label{tab:cuts}
\end{table}

As an example, the left panel of \cref{fig:counts-and-cuts} shows the distributions of the $\tau_\decay$ parameter for the two calibration photon wavelengths (green and magenta) and for the DM search measurement (shaded gray). The horizontal bars labeled ``Tight cut'' and ``Loose cut'' show the cut imposed on $\tau_{\decay}$ in each of these analyses. Any pulse with $\tau_{\decay}$ outside the indicated range is discarded, meaning that the vast majority of pulses identified in the DM search are rejected on this basis alone. The surviving pulses likely cannot be accounted for by noise, even though these dominate our trigger rate during measurement. For confirmation, we have simulated over \qty{500}{\hour} of noise data, from which only one pulse survives the loose cuts and no pulses survive the tight cuts.

Next, we consider the acceptance, or efficiency, of our pipeline for actual energy-deposition events. The total acceptance combines three components: those of the hardware trigger\footnote{Due to the finite energy resolution of our device, the pulse height fluctuates. Therefore, for lower energy pulses with a pulse height closer to the trigger level, the acceptance is energy dependent.} $\alpha^\mathrm{trig}$; the pulse-finding analysis $\alpha^\mathrm{pulse}$; and the cut $\alpha^\mathrm{cuts}$, which takes different values for the loose and tight cuts. The signals simulated with {\tt{TESPASS}} include the corresponding noise baseline, and the trigger mechanism. This allows us to gauge the acceptance of the hardware trigger, which we find to be ${>}\qty{99}{\percent}$ for pulses between \qty{0.6}{\electronvolt} and \qty{3}{\electronvolt}, and degrading for lower energies. The simulated pulses are subsequently analyzed using the pulse-finder again, where the number of pulses surviving the pulse-finder indicates the analysis efficiency. We multiply these efficiencies with the hardware trigger acceptance for each \qty{0.1}{\electronvolt} energy bin. This combined efficiency is then multiplied with the cut acceptance in a next step to determine the total acceptance $\mathcal A_i$ in the $i$th energy bin:
\begin{equation}
     \mathcal A_i = \alpha^\mathrm{trig}_i
        \times\alpha^\mathrm{pulse}_i
        \times\alpha^\mathrm{cuts}_i
        .
\end{equation}
The total acceptance curves for the loose and tight cuts are shown in the top-right panel of \cref{fig:counts-and-cuts}. 

Finally, we evaluate the energy spectrum measured in our DM search using the affine relationship between the energy deposited, $E$, and the pulse integral, $\mathcal I$ (\cref{fig:calib}). Specifically, we take
\begin{equation}
    E = \frac{\mathcal I-\qty{1.76}{\micro\second.\milli\volt}}{\qty{33.53}{\micro\second.\milli\volt.\eV^{-1}}}\,.
\end{equation}
We use the pulse integral parameter rather than the pulse height, even though the pulse height exhibits improved energy resolution, since simulations predict that the affine relationship between pulse integral and deposited energy holds over a larger energy range~\cite{RubieraGimeno:2024}. The resulting spectrum is shown in the bottom-right panel of \cref{fig:counts-and-cuts}.

\subsection{Dark counts and constraint procedure}

Over the course of a \qty{489}{\hour} science run, we identify 126 counts passing the loose cuts and 13 counts passing the tight cuts. The energy distribution of these counts is shown in the lower-right panel of \cref{fig:counts-and-cuts}. We denote these count distributions by $N_i^{\obs}$.

To constrain the DM interaction rate, we use the following procedure. We begin with a signal model that predicts the number of DM events in each bin as a function of the DM mass $m_\dm$, mediator mass $m_\med$, and interaction cross section $\sigma$, which we denote by $N_i^\dm$, where $i$ indexes the energy bins. (The details of the signal model are described in the next section.) We multiply this by the fractional acceptance $\mathcal A_i$ in each bin, shown in the top-right panel of \cref{fig:counts-and-cuts}, to obtain a prediction $N_i^{\mathrm{signal}}(m_\dm, m_\med, \sigma) \equiv \mathcal A_i N_i^\dm$ for the number of observed DM events. We consider only the light- and heavy-mediator limits, i.e., $m_\med \ll q$ or $m_\med \gg q$, where $q$ is the typical momentum transfer to the tungsten.

In order to set a conservative limit, we make no assumption on the nature of the observed events: in principle, they might originate from DM or from uncontrolled backgrounds. The analysis we perform is then not capable of establishing a DM discovery, but only setting a constraint. For each choice of $m_\dm$ and $m_\med$, we compute the maximum value of the interaction cross section $\sigma$ that is compatible with the observed counts allowing any subset of these counts to be attributed to backgrounds. Specifically, we fix the DM mass and mediator mass, and use the profile likelihood ratio test as follows. We assume that the rate of background (non-DM) events in each bin is a Poissonian random variable, with mean $N_i^\bg$. Then the problem is to constrain the DM cross section, $\sigma$, with undetermined nuisance parameters, $N_i^\bg$. For each combination of the model and nuisance parameters, we define
\begin{align}
    &\ell_\dm \equiv \log\mathcal L_{\mathrm{P}}\bigl[
            N_i^\obs \big| N_i^\bg + N_i^{\mathrm{signal}}(\sigma) \bigr],
    \\
    &\ell_\bg \equiv \log\mathcal L_{\mathrm{P}}\bigl[
        N_i^\obs \big| N_i^\bg \bigr],
\end{align}
where $\mathcal L_{\mathrm{P}}(N_i | M_i)$ is the Poisson likelihood of drawing a sample $N_i$ from a multivariate Poisson with mean $M_i$, corresponding to the count rate in each energy bin. We then evaluate the profile likelihood ratio test statistic as
\begin{equation}
    \lambda \equiv 2\left[
        \max_{\{\sigma, N_i^\bg\}}\ell_\dm
        - \max_{\{N_i^\bg\}}\ell_\bg
    \right]
    ,
\end{equation}
maximizing over all possible mean background rate vectors $N_i^\bg$. Note that $\ell_\bg$ is always maximized by taking $N_i^\bg = N_i^\obs$. Since we test each DM mass independently, the signal model has only one parameter, namely the cross section $\sigma$. We thus treat $\lambda$ as a $\chi^2$-distributed random variable with one degree of freedom under Wilks' theorem, and exclude a parameter point if $\lambda$ lies above the 95\% quantile, corresponding to the 95\% confidence level (C.L.). This corresponds to $\lambda > 2.71$ for a one-sided confidence interval.

While this procedure is more computationally complex than a simple counting test, it allows us to take full advantage of the spectral information offered by the energy resolution of the TES, while at the same time imposing no assumptions on the backgrounds in the experiment. The only remaining input for setting constraints on DM interactions is the signal model itself, which we describe in the next section.

\section{Dark Matter Interaction Rate}\label{sec:DM}

Our TES device is sensitive to several different channels of DM interactions, including \textbf{(1)} DM-electron scattering, \textbf{(2)} DM absorption onto electrons, and \textbf{(3)} DM-nucleon scattering. In each case, we consider the interaction occuring in the TES sensor. When depositing energy above the effective threshold of the device, such an interaction would result in a voltage pulse which can be read out. As long as the energy deposited by the DM is not driving it into the normal conducting state and happening on a fast time scale, it functions as a linear calorimeter to deposited energy, regardless of the energy’s source, meaning that the TES behaves identically in a DM interaction event and a photon absorption event. (See \refscite{Hochberg:2021ymx,Griffin:2024cew} for further details.) As opposed to other superconducting detectors like SNSPDs, the device’s effective energy threshold is not defined by a distinct energy, but rather the system’s noise level. Thus, by evaluating the rates of these processes as a function of DM parameters, we can use the observed DM search data rate to constrain the DM parameter space.

We compute the rate for each of  the above processes following \refscite{Hochberg:2021pkt,Hochberg:2021yud,Griffin:2024cew}, assuming that the $\dm$ interacts with Standard Model species via a mediator $\med$. The event rate per unit detector mass is given by 
\begin{multline}
    \Gamma =
        \frac{\pi n_\dm\bar\sigma_t}{\mu_{t,\dm}^2}
        \int\frac{\du^3\bb v_\dm\dd^3\bb q\dd\omega}{(2\pi)^3}
        \,f_\dm(\bb v_\dm)
        \\
        \times\mathcal F(q)^2\, S(\bb q, \omega)
        \delta(\omega - \omega_{\bb q})\,
    ,
\end{multline}
where $n_\dm$ is the DM number density; $\bar\sigma_t$ is a reference cross section for DM-$t$ scattering, with $t$ denoting the target particle; $\mu_{t,\dm}$ is the reduced mass; $\omega$ is the energy deposited; $\bb q$ is the 3-momentum transfer; $\bb v_\dm$ is the DM velocity; $\omega_{\bb q} = \bb q\cdot\bb v_\dm - \bb q^2/2m_\dm$ is the energy deposited in the detector; $f_\dm(\bb v_\dm)$ is the DM velocity distribution function in the laboratory frame; $\mathcal F(q)$ is a model-dependent form factor; and $S(\bb q, \omega)$ is the dynamic structure factor, given by Fermi's Golden Rule. For $f_\dm(\bb v_\dm)$ we assume the standard halo model~\cite{Lewin:1995rx}. We take the local DM density to \qty{0.4}{\giga\electronvolt/\centi\meter^3}, the DM velocity dispersion to be \qty{220}{\kilo\meter/\second}, the escape velocity to be \qty{550}{\kilo\meter/\second}, and Earth velocity \qty{232}{\kilo\meter/\second}, with all velocities specified in the Galactic frame. For scattering via a spin-independent interaction, the form factor is given by $\mathcal F(q) = (m_\phi^2+q_{0,t}^2)/(m_\phi^2+q^2)$, where $m_\med$ is the mass of mediator, and $q_{0,t}$ is a reference momentum. We take the reference momentum for electronic scattering to be $q_{0,e} \equiv \alpha\, m_e$, with $\alpha$ the fine structure constant, and we take $q_{0,n} \equiv m_\dm\langle v_\dm\rangle$ for nuclear scattering. In both cases, the reference cross section is given by $\bar\sigma_t \equiv \frac1\pi\mu_{t,\dm}^2g_0^2/(m_\med^2 + q_{0,t}^2)^2$, where $g_0$ stands in for coupling constants. Given the dynamic structure factor for each DM interaction channel of interest, one can now compute the corresponding event rate.

For DM interactions with electrons---both scattering and absorption---we use the linear response theory of dielectric systems~\cite{Hochberg:2021pkt}. For spin-independent scattering with electrons, the dynamic structure factor is given by 
\begin{equation}\label{eq:Sscat}
    S(\bb q, \omega) = \frac{2 \bb q^2}{e^2}\Im\left(\frac{-1}{\epsilon(\bb q, \omega)}\right)
\end{equation}
with $\epsilon(\bb q, \omega)$ denoting the dielectric function of the material. For DM absorption, we consider the case of a kinetically-mixed dark photon DM, where the interaction Lagrangian has the form $\mathcal L_{\mathrm{int}} = -\frac12\kappa F_{\mu\nu}F^{\prime\mu\nu}$. Here $\kappa$ is a coupling constant, and $F_{\mu\nu}$ is the field strength tensor, \textit{i.e.,} $F_{\mu\nu} \equiv \partial_\mu A_\nu - \partial_\nu A_\mu$, where $A_\mu$ is the photon field. Primes refer to the dark photon field $A'_\mu$ in place of the photon field. The absorption rate in this case is given by~\cite{Hochberg:2021yud}
\begin{equation}
    \Gamma_{\mathrm{A}} =\kappa^2m_\dm
        \Im\left(-\frac{1}{\epsilon(m_\dm\bb v_\dm,\,m_\dm)}\right)\,.
\end{equation} 
Since $|\bb v_\dm|\sim\num{e-3}$, we have $m_\dm|\bb v_\dm| \ll m_\dm$, and so in practice absorption is determined by the $\bb q \rightarrow 0$ limit of the dielectric function. Note that the dynamic structure factor can receive corrections from geometry for thin-layer detectors when the momentum transfer $|\bb q|$ is not large compared to the inverse layer thickness~\cite{Lasenby:2021wsc,Hochberg:2021yud,QROCODILE:2024zmg}. In our experiment, the inverse thickness of the superconducting layer is $1/(\qty{20}{\nano\meter}) \approx \qty{10}{\electronvolt}$, whereas typical momentum transfers are of order $v_\dm m_\dm \gtrsim \num{e-3}\times\qty{30}{\kilo\electronvolt} = \qty{30}{\electronvolt}$ for all DM masses we consider in this work. Thus, we neglect geometric corrections in evaluating the DM interaction rate.

Our TES sensor is also sensitive to the scattering of DM particles with nuclei, as it can be triggered by a nuclear scattering event in the device via phonon production, as detailed by \refcite{Griffin:2024cew}. Here we place a conservative limit on DM interactions with nuclei via nuclear recoils, where the dynamic structure factor is given by~\cite{Trickle:2019nya}:
\begin{equation}
    \label{eq:structure-factor-elastic}
    S(\bb q, \omega) =
    \frac{2\pi\rho_{\mathrm{T}}}{\sum_{N} A_{N}} \sum_{N} \frac{A_N^3}{m_N}
    F_N(\bb q)^2
    \delta \left(\omega - \frac{\bb q^2}{2 m_N}\right)
    \,.
\end{equation}
In the above, $N$ indexes the nuclei in a unit cell; $m_N$ is the atomic mass; $A_N = m_N/\qty{}{u}$ is the atomic mass number; $f_n$ is the coupling to DM; and $F_N(\bb q)$ is the nuclear form factor. We take the Helm form factor~\cite{Helm:1956zz}, $F_N (q) = [3 j_1 (q r_N)/(q r_N)] e^{-(qs)^2/2}$,  with $q=|{\bb q}|$, $j_1$ the spherical Bessel function of the first kind, $r_N \approx A_N^{1/3}\times\qty{1.14}{\femto\meter}$  the effective nuclear radius, and $s$ the nuclear skin thickness. We use $A_{\ce{W}}\approx183.85$ and $s = \qty{0.9}{\femto\meter}$. In principle, the low effective threshold of our device would allow sensitivity to even lower DM masses via multiphonon production. The reach is then determined by the vibrational spectrum for the tungsten used in our detector. The particular composition of the tungsten in our TES requires a dedicated study of its vibrational spectrum, and we thus relegate such an analysis to future work.

\begin{figure*}
\centering
    \includegraphics[width=\textwidth]{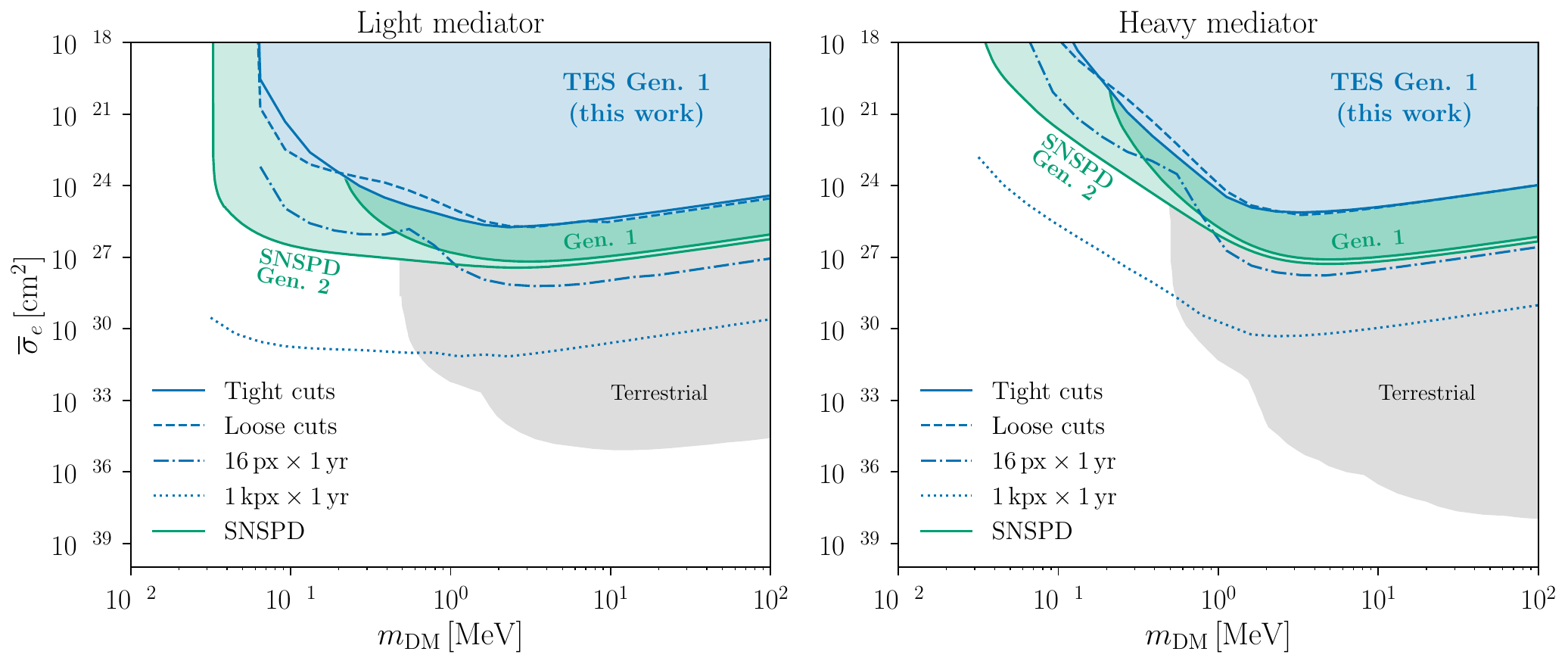}
    \caption{DM-electron scattering. New bound and projections at the 95\% C.L. on DM scattering with electrons via a light~(\textit{left}) or heavy~(\textit{right}) mediator in a TES sensor. In both panels, the shaded blue region indicates the new constraint we place with the existing TES data from DESY using the tight (solid curve) or loose (dashed curve) cuts. The projected reach of a future experiment using a 16-pixel array exposed for a year and using the loose cuts, with each unit similar to our current TES and with background scaling with exposure, is shown in dot-dashed curve. The projection for a 1000 pixel array exposed for a year, assuming a lower energy threshold of \qty{70}{\milli\electronvolt} and  no backgrounds is shown in dotted blue. For comparison, we show existing constraints from SNSPD sensors~\cite{Hochberg:2019cyy,Hochberg:2021yud,QROCODILE:2024zmg}, indicating the improvement achieved from first to second generation devices. Other existing terrestrial constraints are shown in shaded gray~\cite{Barak:2020fql, Amaral:2020ryn, Aguilar-Arevalo:2019wdi, Essig:2017kqs, Agnes:2018oej, XENON:2019gfn}. Constraints may be relaxed for $\overline{\sigma}_e\gtrsim\qty{e-24}{\centi\meter^2}$ due to atmospheric scattering (see text).
    }
    \label{fig:e-scattering}
\end{figure*}

\section{Results}\label{sec:res}

\Cref{fig:e-scattering} shows the constraints we derive at 95\% C.L. on DM-electron scattering using DM search data from the ALPS~II TES sensor, for the case of a light~(left) or heavy mediator~(right). In each panel, the new constraint is shown by the blue shaded region, bounded by the stronger of the results from the tight (solid) and loose (dashed) cuts. We also show projections for future iterations of this experiment. The dot-dashed curve shows the anticipated limit for a 16-pixel TES array whose units have the same parameters as our device, using the loose cuts and operated for one year. This curve assumes that the number of observed counts scales directly with the exposure. The dotted curve shows the sensitivity of a 1000-pixel array exposed for a year. Here, we assume that the pixels have the same volume and composition as our device, but operate with an effective threshold of \qty{70}{\milli\electronvolt}, roughly corresponding to the energy resolution of the device demonstrated in \refcite{Hattori:2022mze}. We further assume that the experiment observes zero counts, \textit{i.e.}, the most optimistic circumstances for DM constraints.

For comparison, we also show the constraints placed by first-~\cite{Hochberg:2019cyy,Hochberg:2021yud} and second-generation~\cite{QROCODILE:2024zmg} SNSPD DM searches in shaded green. The threshold of our first-generation TES is already lower than that of first-generation SNSPD devices, enabling sensitivity to lower DM masses. The impressive extended reach of the second-generation SNSPD device compared to a first-generation sensor demonstrates the great strides that can be made in a short time with dedicated R\&D efforts focused on optimizing the technology for light DM searches.

\begin{figure}
    \centering
    \includegraphics[width=0.5\textwidth]{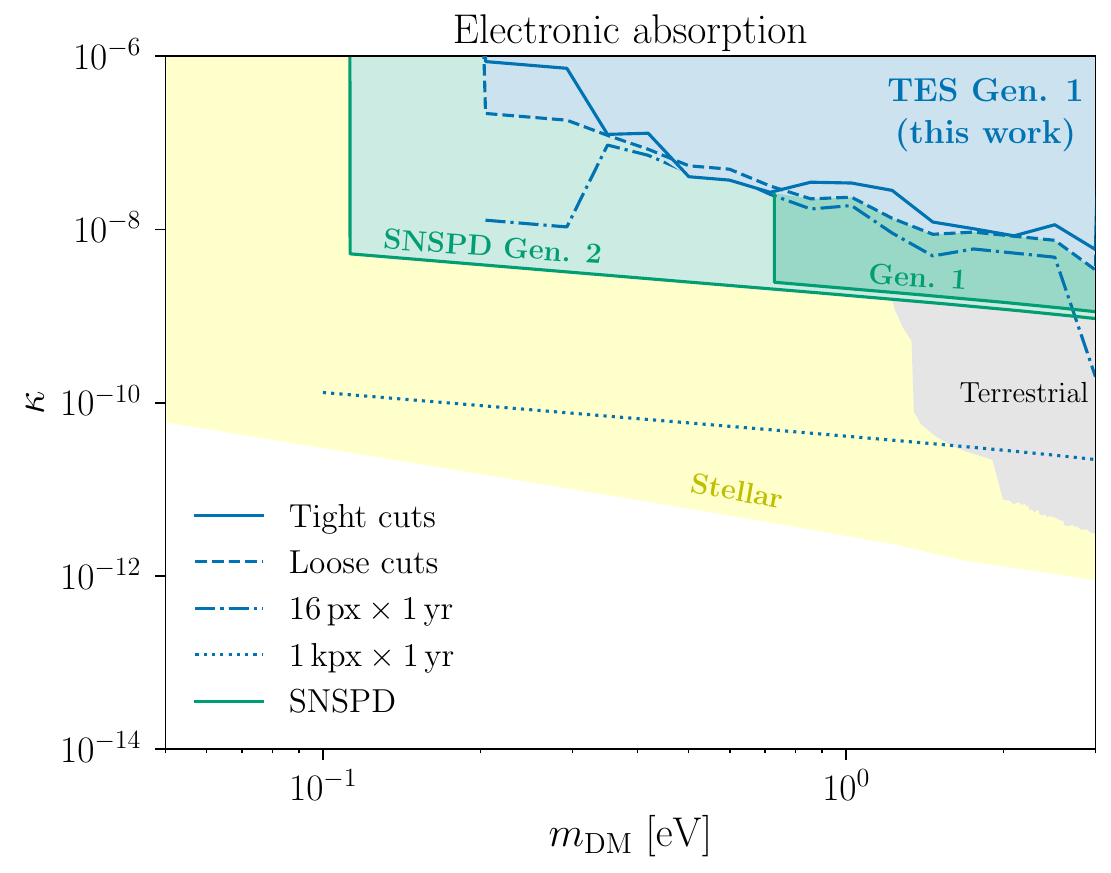}
    \caption{Dark photon DM absorption. New TES limit and projections at the 95\% C.L. on dark photon DM absorption on electrons. Model-dependent stellar constraints~\cite{An:2013yua, An:2014twa, An:2020bxd} are shown in shaded yellow. All other features are identical to those of \cref{fig:e-scattering}.
    Constraints may be relaxed due to atmospheric overburden (see text).
    }
    \label{fig:absorption}
\end{figure}

\begin{figure*}
    \centering
    \includegraphics[width=\textwidth]{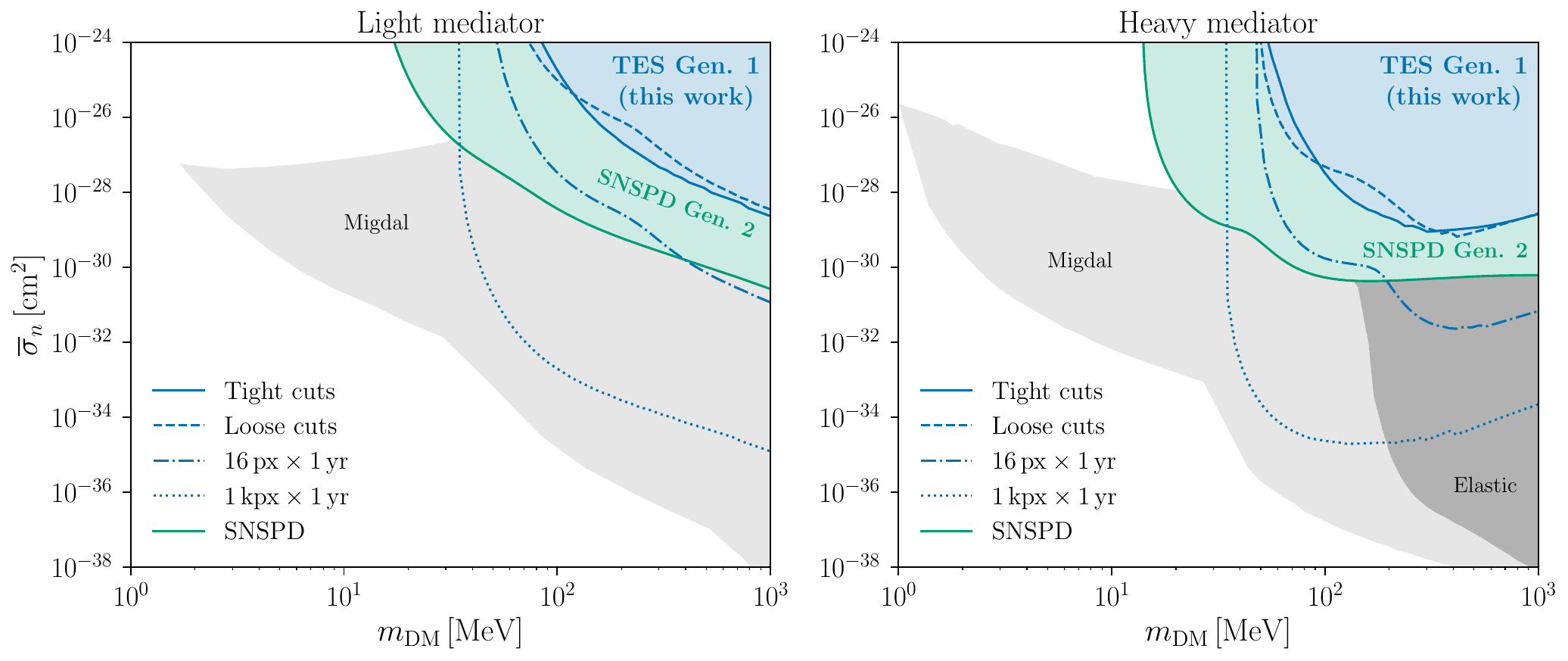}
    \caption{DM-nucleon scattering. New limits and projections at the 95\% C.L. for DM-nucleon scattering. The shaded dark and light gray regions indicate the strongest existing terrestrial constraints to date~\cite{EDELWEISS:2019vjv, EDELWEISS:2022ktt,DarkSide-50:2022qzh, Franco:2023sjx,SuperCDMS:2020aus, SuperCDMS:2023sql,CRESST:2019jnq,SENSEI:2023zdf,PandaX:2023xgl,LUX:2018akb,Essig:2019xkx} based on elastic interactions and the Migdal effect, respectively.  
    All other features are identical to those of \cref{fig:e-scattering}. Constraints may be relaxed for $\overline{\sigma}_n\gtrsim\qty{e-25}{\centi\meter^2}$ due to atmospheric scattering (see text).}
    \label{fig:n-scattering}
\end{figure*}

\Cref{fig:absorption} shows limits on kinetically-mixed dark photon DM from absorption in the TES, with the same conventions as \cref{fig:e-scattering}. First-~\cite{Hochberg:2019cyy,Hochberg:2021yud} and second-generation~\cite{QROCODILE:2024zmg} SNSPD constraints are again shown in shaded green, with other terrestrial constraints~\cite{An:2014twa, Agnese:2018col, Aguilar-Arevalo:2019wdi, Arnaud:2020svb, FUNKExperiment:2020ofv, Barak:2020fql} indicated in shaded gray. Model-dependent complementary stellar constraints~\cite{An:2013yua, An:2014twa, An:2020bxd} appear in shaded yellow. Note that the impact of backgrounds is much more sharply mass-dependent for absorption than for scattering. This is because the predicted spectrum for absorption is a narrow spike at an energy corresponding to the DM mass, meaning that the background for an absorption search at a given mass is simply the count rate in that bin. The high observed count rate for $\qty{0.5}{\electronvolt}<\omega<\qty{1}{\electronvolt}$ (\cref{fig:counts-and-cuts}) accounts for the weaker scaling of the projected 16-px limit in this regime. For scattering, on the other hand, the fact that the spectral shape of the observed background is a poor fit to the DM prediction allows for substantial background mitigation.

Our new constraints and projections on light DM scattering with nucleons are similarly depicted in \cref{fig:n-scattering}, along with existing limits from elastic DM-nucleon scattering and the Migdal effect~\cite{Ibe:2017yqa}. The tungsten film of our TES sensor has a polycrystalline structure consisting of both $\alpha$- and $\beta$-phase tungsten, meaning that the computation of the vibrational spectrum is nontrivial and will be the subject of future work. In the present work, for DM-nucleon interactions, we conservatively place limits and show projections only using nuclear recoils in the TES rather than considering single-phonon excitations, as described in \refscite{Campbell-Deem:2022fqm,Griffin:2024cew}. Significant improvement should be possible at low DM masses once the phonon spectrum of the device is established.

Note that the cross sections we show for DM-electron and DM-nucleon scattering towards the top of \cref{fig:e-scattering,fig:n-scattering} sit above where atmospheric scattering may occur, and overburden effects may become significant. The quantitative inclusion of overburden is nontrivial and model-dependent, but the constraints become unreliable at cross sections above ${\sim}\qty{e-24}{\centi\meter^2}$ (${\sim}\qty{e-25}{\centi\meter^2}$) for DM-electron (DM-nucleon) scattering. (See \refscite{Emken:2017erx,Emken:2017qmp,Emken:2018run,Emken:2019tni} for details. Note that these works do not compute atmospheric overburden for electronic absorption, potentially relevant to \cref{fig:absorption}. A full treatment of atmospheric absorption is beyond the scope of this work.) Nonetheless, the importance of our work is in demonstrating the reach of a current TES technology as a first-generation DM detector, and to motivate future experimental configurations that scale up in exposure, push energy thresholds lower, and keep noise down, which will enable sensitivity to interesting new regions of light DM parameter space. Indeed, SNSPD technologies have demonstrated that remarkable advances can be achieved between first- and second-generation devices, and there is no fundamental limitation to similar headway being made with future TES designs, especially considering their energy-resolving capabilities.

\section{Outlook}
\label{sec:out}

We have proposed and prototyped the use of TES detectors as simultaneous targets and sensors for light DM detection with sub-MeV mass. We use an existing TES detector system, intended for use as a single-photon detector for axion-like particle searches in the ALPS~II experiment, as a proof of concept. In a setup similar to the one used for the ALPS~II experiment, the TES is  directly sensitive to sub-MeV DM while operating in a dilution refrigerator and after removing optical fibers used for photon detection, and we have improved sensitivity at low masses by reducing the effective energy threshold of the TES. Future configurations that increase exposure, further reduce the energy threshold, and preserve the low dark count rate should enable a dedicated TES experiment to probe untested regions of light DM parameter space through several channels, including DM-electron scattering, DM-nucleon scattering, and DM absorption.

The advancement of quantum sensors opens exciting possibilities for probing light DM. SNSPDs are one such technology that has gained much traction in the context of sub-MeV DM detection, with the dedicated QROCODILE experimental collaboration advancing SNSPD sensors in size, threshold and exposure. Both SNSPDs and TESs have demonstrated low sub-eV energy thresholds and low dark count rates, and along with KIDs offer a pathway towards significant progress in light DM detection.  The strength of TESs comes from their inherent energy resolution: unlike SNSPDs which are counting devices, the signal pulses in a TES contain the energy information of the incident particle, which is proportional to the integral of the pulse area. In previous measurements, our TES showed a fiber-coupled energy resolution of $(5.31\pm 0.06)\%$ at \qty{1.165}{\electronvolt}~\cite{RubieraGimeno:2023J9}. Efforts to improve this energy resolution further include tuning the structure of the TES films toward reducing the critical temperature $T_\mathrm{C}$~\cite{Lita2005,Hattori:2022mze, 2017ApPhL.110u2602M}. The measured energy spectrum can be used in the analysis to improve limits, as we demonstrate here, and will also be crucial to identifying and scrutinizing any putative DM signal. 

Improving the sensitivity to small cross sections will require a larger fiducial volume in future iterations of this experiment. In \cref{fig:absorption,fig:e-scattering,fig:n-scattering}, we have shown projections for experiments with 16 and \num{1000} pixels, respectively. The choice of a 16-pixel array is motivated by existing design proposals at a similar scale~\cite{Szypryt:LTD2025}. Future extensions of readout schemes currently being developed~\cite{Szypryt:2024gbp} are anticipated to enable kilopixel-scale arrays in the coming few years. As such, the projections we show are realistic estimates of progress that may be made by a second-generation experiment in the next few years.

With further optimization of TES detectors dedicated for DM searches, much improvement to the reach of TES sensors into light DM parameter space can be achieved. Reaching the simplest cosmological target model in this mass range, corresponding to freeze-in production of DM~\cite{Hall:2009bx,Elahi:2014fsa}, would require an improvement of many orders of magnitude~\cite{Essig:2015cda,Dvorkin:2019zdi} to our current sensitivity. However, several other DM production mechanisms are viable in this mass range (see e.g.~\refscite{Zurek:2013wia,Hochberg:2014dra,Hochberg:2014kqa,Kuflik:2015isi,Kuflik:2017iqs}), and there is thus no single prediction or even expectation for the DM cross section. We may therefore discover DM interactions at cross sections much higher than those corresponding to freeze-in production.

In the short term, a number of improvements are possible. At the level of the analysis, optimization of cuts and machine learning techniques~\cite{Meyer:2023ffd} may enhance sensitivity. To improve the distinguishability of backgrounds and possible events in future configurations, one can leverage TES modules that have been designed specifically for direct DM searches with possible veto sensors (such as SNSPDs) or substrate readout. One version of an optimized sensor---without zirconia sleeves for optical fiber insertion---is currently being tested in the TES laboratory at DESY. Removal of the zirconia sleeves may reduce the rate of initial triggers at low trigger thresholds from radioactive backgrounds. Another TES placed on a \ce{SiN_x} membrane instead of the regular substrate is currently being assembled at PTB Berlin. By reducing the active material around the TES, we expect to reduce backgrounds from energy deposits in the substrate.

Given the prospects when combining such optimized sensors with scaled arrays, reduced thresholds, and extended exposures, TES technology promises to offer a powerful new probe of light DM parameter space. This work thus paves the way for a new class of experiments to lead the light DM searches in the coming years.

\bigskip

\textit{Note added.} During the final stages of completion of this manuscript, \refcite{Chen:2025cvl} appeared, which also discusses the use of TES sensors for light DM detection, in similar spirit to our previous works of \refscite{Schwemmbauer:2024rcr,Schwemmbauer:2024jel}.

\bigskip

\textit{In Memoriam.} We regretfully acknowledge the untimely passing of Sae Woo Nam, who collaborated with us during early stages of this project. We dedicate this article to his memory.

\bigskip

\begin{acknowledgments}
We thank Jörn Beyer from PTB Berlin as well as Marco Schmidt from Humboldt-Universität zu Berlin for vital advice and support. The work of Y.H.\ is supported in part by the Israel Science Foundation (grant No.\ 1818/22) and by the Binational Science Foundation (grants No.\ 2018140 and No.\ 2022287). The work of Y.H.\ and G.D.H.\ is supported by an ERC STG grant (``Light-Dark,'' grant No.\ 101040019). The work of B.V.L.\ is supported by the MIT Pappalardo Fellowship.  The work of A.E.L.\ is supported in part by NIST. M.M.\ and E.R.\ acknowledge the European Research Council (ERC) support under the European Union’s Horizon 2020 research and innovation program Grant agreement No. 948689 (AxionDM). C.S., F.J., A.L.\ and M.M.\ acknowledge the support by the Deutsche Forschungsgemeinschaft (DFG, German Research Foundation) under Germany’s Excellence Strategy -- EXC 2121 ``Quantum Universe'' -- 390833306. This work was supported via the project Quantum Sensing for Fundamental Physics (QS4Physics) from the Innovation pool of the research field Helmholtz Matter of the Helmholtz Association. This project has received funding from the European Research Council (ERC) under the European Union’s Horizon Europe research and innovation programme (grant agreement No.\ 101040019).  Views and opinions expressed are however those of the author(s) only and do not necessarily reflect those of the European Union. The European Union cannot be held responsible for them. Certain commercial equipment, instruments, or materials are identified in this paper to foster understanding. Such identification does not imply recommendation or endorsement by the National Institute of Standards and Technology, nor does it imply that the materials or equipment identified are necessarily the best available for the purpose.
\end{acknowledgments}

\bibliography{references}

\end{document}